%% file: main.tex
\documentclass[sigconf]{acmart}

\usepackage{fontawesome}
\usepackage{multirow}
\definecolor{Gallery}{rgb}{0.937,0.937,0.937}
\usepackage{colortbl}
\usepackage{soul}
\usepackage{enumitem}
\usepackage{fp}

\definecolor{darkgreen}{rgb}{0.0, 0.6, 0.2} 
\definecolor{lightgreen}{rgb}{0.3, 0.5, 0.4} 

\AtBeginDocument{%
  }

\copyrightyear{2026}
\acmYear{2026}
\setcopyright{cc}
\setcctype{by}
\acmConference[ICSE-SEIS '26]{2026 IEEE/ACM 48th International Conference on Software Engineering}{April 12--18, 2026}{Rio de Janeiro, Brazil}
\acmBooktitle{2026 IEEE/ACM 48th International Conference on Software Engineering (ICSE-SEIS '26), April 12--18, 2026, Rio de Janeiro, Brazil}
\acmPrice{}
\acmDOI{10.1145/3786581.3786934}
\acmISBN{979-8-4007-2424-4/2026/04}

\begin{document}

\def\barChart#1{%
  \begingroup
  \FPmul\BarLen{#1}{0.015}%
  \raisebox{0.1ex}{%
    \hbox{%
      {\color[rgb]{0.30,0.60,0.45}\rule{\BarLen cm}{1.2em}}%
      \hspace{0.4em}%
      {\normalsize\bfseries #1\%}%
    }%
  }%
  \endgroup
}



\makeatletter
\def\miniHistogram{\@ifnextchar[{\miniHistogram@opt}{\miniHistogram@opt[5]}}%

\def\miniHistogram@opt[#1]#2{%
  \leavevmode
  \vbox{%
    \hbox{%
      \def\maxval{#1}%
      \def\barwidth{6pt}
      \def\totalH{15}
      \def\do##1{%
        \FPeval\h{(##1)/(\maxval)*(\totalH)}%
        {\color{black!70}\rule{\barwidth}{\h pt}}\hspace{0.5pt}%
      }%
      \forcsvlist{\do}{#2}%
    }%
    \hrule height 0.8pt%
  }%
}
\makeatother

\title{From Gains to Strains: Modeling Developer Burnout with GenAI Adoption} 

\author{Zixuan Feng}
\affiliation{%
  \institution{Oregon State University}
  \city{Corvallis}
  \state{Oregon}
  \country{USA}
}
\email{fengzi@oregonstate.edu}

\author{Sadia Afroz}
\authornote{Co-first author}
\affiliation{%
  \institution{Oregon State University}
  \city{Corvallis}
  \state{Oregon}
  \country{USA}
}
\email{afrozs@oregonstate.edu}

\author{Anita Sarma}
\affiliation{%
  \institution{Oregon State University}
  \city{Corvallis}
  \state{Oregon}
  \country{USA}
}
\email{anita.sarma@oregonstate.edu}

\begin{abstract}

Generative AI (GenAI) is rapidly reshaping software development workflows. While prior studies emphasize productivity gains, the adoption of GenAI also introduces new pressures that may harm developers’ well-being. In this paper, we investigate the relationship between the adoption of GenAI and developers' burnout. We utilized the Job Demands–Resources (JD--R) model as the analytic lens in our empirical study. We employed a concurrent embedded mixed-methods research design, integrating quantitative and qualitative evidence. We first surveyed 442 developers across diverse organizations, roles, and levels of experience. We then employed Partial Least Squares–Structural Equation Modeling (PLS-SEM) and regression to model the relationships among job demands, job resources, and burnout, complemented by a qualitative analysis of open-ended responses to contextualize the quantitative findings. Our results show that GenAI adoption heightens burnout by increasing job demands, while job resources and positive perceptions of GenAI mitigate these effects, reframing adoption as an opportunity.

\end{abstract}

%
%
%

\ccsdesc[300]{Human-centered computing~Empirical studies in collaborative and social computing}

\ccsdesc[100]{Human-centered computing}
\keywords{Generative AI, Developer Burnout, Job Demands–Resources (JD–R)}

\maketitle

\input{sections/1_intro}

\label{sec:intro}

\input{sections/2_related_work}

\label{sec:related}

\input{sections/2.5_Theory}

\label{sec:theory}

\input{sections/3_Survey}

\label{sec:survey}

\input{sections/4_RQ1_PLS}

\input{sections/5_RQ2_logistic}

\input{sections/6_limitation}
\label{sec:limitation}

\input{sections/7_discussion}

\label{sec:discussion}

\begin{acks}
We thank all survey participants for their time and insights. This work was supported by NSF Grant No. 2303043.
\end{acks}

\bibliographystyle{ACM-Reference-Format}
\bibliography{bib}
\end{document}

%% file: sections/1_intro.tex
\section{Introduction}
We are now at a pivotal shift in how Generative AI (GenAI) is transforming how we work and live. Since late 2022, GenAI has progressed from experimental pilots to mainstream adoption, reshaping how code is written, reviewed, and maintained \cite{nguyen2025generative, oliverwyman2024ai}. Individual adoption is rapid---84\% of developers in the 2025 Stack Overflow survey reported using or planning to use GenAI tools \cite{StackOverflow2025_AItools}. Organizations are not far behind. Motivated by promises of higher productivity, they are heavily investing in GenAI \cite{GitHub2025_CopilotImpact}.

Some studies support these investments. For example, GitHub reported that developers completed coding tasks 55.8\% faster \cite{GitHub2025_CopilotImpact}. Yet, others report increased costs. For instance, the 2025 State of Software Delivery Report by Harness \cite{harness2025} reported 67\% of developers spent more time debugging AI-generated code and 68\% spent more time fixing AI-created security issues. Similarly, \citet{becker2025measuring} reported an overall 19\% productivity loss in their observation study. These contrasting data suggest that efficiency gains from GenAI are not uniform, often redistributing development effort from content creation to verification and oversight  \cite{dora2024, harness2025, StackOverflow2025_AItools}.

Organizations, however, are adopting genAI at a rapid pace, and in the process escalating developer productivity expectations and job automations \cite{HackerRank2025_PressureToDeliver}. Labor markets reflect these trends: the Stanford Digital Economy Lab reported entry-level cognitive jobs, disproportionately held by early-career developers, have declined by 13–20\% \cite{brynjolfsson2025canaries, bernard2025tech}. Other reports also show reduced headcounts because of AI automation \cite{Cerullo2025_CBSjobsLayoffs, WEF2025_AIJobsLabourDay}. Automation has another hidden cost on the workforce; it causes reduced situational awareness and decision-making skills, leading to actual \cite{kaber1997out} and perceived fear of deskilling \cite{crowston2025deskilling}. To avoid such deskilling and to keep up with the rapid pace of AI development, developers have to continually adapt and reskill to stay relevant \cite{qiu2025today, necula2023artificial}.

Burnout---characterized by a chronic state of exhaustion, emotional distance, cynicism, and diminished professional efficacy \cite{schaufeli2004job, lazarus1985stress, anand2023effect}---is a foreseeable outcome in this situation. 

Burnout itself is not new to Software Engineering (SE); it has surfaced during prior disruptions, from the rise of agile \cite{tulili2023burnout} to the COVID-19 pandemic \cite{cao2024jd}. Past works, however, have not theorized the impact of genAI adoption on burnout. Instead, these studies have largely focused on developer productivity (e.g., speedups, throughput) \cite{houck2025space, becker2025measuring}, creativity \cite{jackson2024creativity}, or adoption intent \cite{russo2024navigating, johnson2023make, choudhuri2025needs}, without researching its effects on developers' well-being. 

Examining the effects of GenAI adoption on developer well-being is necessary to ensure that the promised productivity gains do not come at the cost of developer health and the sustainability, particularly as work demands and access to resources are seldom distributed evenly across role, seniority, or organization context.
In this work, we therefore ask: 

\begin{itemize}[leftmargin=*, label={}, nosep]
    \item \textbf{RQ1:} How is GenAI adoption associated with developer burnout?
    \item \textbf{RQ2:} How do developers' characteristics influence their perceptions of burnout?
\end{itemize}

We answer these questions with a concurrent embedded study \cite{greene2007mixed, morse2016mixed} of a survey of professional developers ($N=442$), integrating quantitative modeling with qualitative evidence through the lens of the Job Demands–Resources (JD--R) model. The JD--R model is particularly well-suited for our context because it conceptualizes burnout as the outcome of an imbalance between job demands (e.g., organizational pressure, workload) and job resources (e.g., autonomy, learning resources), and has been applied in investigating burnout in other domains, such as the impact of COVID-19 \cite{cao2024jd}.

More specifically, we answer \textbf{RQ1} by building a theoretical model using Partial Least Squares--Structural Equation Modeling (PLS-SEM) to estimate the relationships among GenAI adoption, job demands, job resources, and burnout. We answer \textbf{RQ2} through regression analyses that relate developers' characteristics (e.g., role, organization size, industry seniority) to JD--R perceptions and burnout. Following the concurrent embedded design, we qualitatively analyzed open-ended responses to contextualize mechanisms and support quantitative findings.

Our contributions are threefold: (1) We establish that GenAI adoption is linked to burnout through elevated organizational pressure and workload, while autonomy and learning resources, which are unevenly distributed, mitigate these effects; (2) We offer the first explanatory model of when GenAI acts as a demand amplifier versus a resource enabler in software development by extending the JD--R model to AI adoption; (3) We translate our findings into implications for workload design, workforce development, and team practices to help organizations enable equitable workload and resources for their employees.

Our contributions can help organizations leverage the productivity potential of GenAI, without sacrificing developer well-being, as one participant cautioned: \textit{``I move fast with AI and move mountains of work, but I am losing my passion for the work''} [P212] \footnote{P:Survey Participant}.

%% file: sections/2_related_work.tex
\section{Background}

%
GenAI-powered tools are being rapidly integrated into software development workflows \cite{nguyen2025generative, usage2025, GitHub2025_CopilotImpact}, fueled by the \textbf{promise} of productivity gains \cite{qiao2025comprehension, shihab2025effects}. For example, developers have reported task speed-ups ranging from 21\% to 55\% when using GenAI tools \cite{peng2023impact, paradis2025much, kalliamvakou2022copilot}. GitHub Copilot, for example, has been credited with cutting development lead times by more than half \cite{kalliamvakou2022copilot}. \citet{houck2025space} utilized the SPACE framework to demonstrate that AI significantly enhances productivity, particularly for routine tasks. 

These results provide a powerful narrative for organizations to actively encourage developers to integrate GenAI tools, so as to accelerate efficiency and remain competitive \cite{li2024ai}. A growing body of research is investigating how GenAI can be effectively integrated into software development workflows. For example, \citet{esposito2025generative} reviewed 46 studies from over 1,000 screened papers, focusing on how GenAI is applied in software architecture, including its potential to support design decisions. \citet{he2025llm} synthesized 71 studies on LLM-based multi-agent systems, outlining integration pathways across requirements engineering, code generation, quality assurance, and maintenance.

Studies have also investigated the different 
human factors that influence GenAI adoption \cite{afroogh2024trust}. For example, \citet{huynh2025generative} emphasized that fairness, social presence, and emotions are determinants of trust in GenAI, which in turn shapes user attitudes, perceived performance, and adoption intentions. \citet{choudhuri2025needs} demonstrated that developers’ trust and adoption intentions are shaped not only by system quality and functional value but also by cognitive diversity.

But these positive narratives are not uniform, and there are \textbf{perils} to the GenAI hype  \cite{kang2024quantitative, StackOverflow2025_AItools}. A survey of 39,000 developers found only a 2.1\% rise in overall productivity and a 3.4\% improvement in code quality for every 25\% increase in AI adoption, while software delivery performance actually declined by 7.2\% \cite{dora2024}. \citet{becker2025measuring} found that AI tools slowed developers down, increasing task completion time by 19\%.

Beyond efficiency, GenAI adoption introduces new job demands: on one hand, developers face organizational pressure to quickly adopt AI tools; on the other hand, once adopted, they are expected to work at a faster pace \cite{google2025state, jaworski2023study}. At the same time, developers must invest substantial effort in validating, debugging, and securing AI-generated outputs, which often displaces the time saved on initial coding tasks \cite{harness2025, StackOverflow2025_AItools}. However, amid this hype and intensified workload, developers report receiving insufficient training and organizational support \cite{mayer2025superagency, gao2025ai}.

%
Together, these emerging observations suggest that declining performance, rising organizational pressure, heavier workload, and insufficient resources are creating a growing strain for developers. 
Yet, the well-being of developers impacted by these changes remains largely unexplored within the SE society.

%
To address this gap, we turn to the construct of \textbf{burnout}, which captures how work-related stress translates into exhaustion, detachment, and diminished efficacy \cite{maslach2001job}. Burnout can manifest through multiple interrelated aspects of work life, such as feeling drained by demands, developing detached or negative attitudes toward work \cite{schaufeli2004job}, and experiencing uncertainty about job security \cite{anand2023effect}. Several theories could be applied to analyze burnout, including the Transactional Model of Stress \cite{lazarus1985stress}, the Conservation of Resources theory \cite{hobfoll1989conservation}, Social Cognitive Theory \cite{bandura1991social}, and Job Demands–Resources (JD--R) model \cite{demerouti2001job, bakker2017job}. We use the JD--R model as our analytical lens. 

The JD--R model \cite{demerouti2001job, bakker2017job}, developed initially as a general model of burnout \cite{demerouti2001job}, conceptualizes how job demands (e.g., workload, pressure, uncertainty) and job resources (e.g., autonomy, learning opportunities) jointly shape well-being. Its flexibility has made it a widely adopted framework for investigating well-being across diverse work contexts, including software development \cite{bakker2017job, trinkenreich2024predicting}.

%% file: sections/2.5_Theory.tex
\section{Theory Development}

This section explains how we developed our theoretical model by adopting the JD--R perspective in the context of GenAI adoption in software development.

\textbf{Job Demands} refer to the physical, social, or organizational aspects of work that require sustained cognitive effort and are associated with physiological and psychological costs, such as exhaustion \cite{demerouti2001job, bakker2017job}. Foundational JD--R studies \cite{balducci2011job, bakker2007job} conceptualize Job Demands as including key factors like (1) Organizational Pressure and (2) Workload.

\textit{Organizational pressure}. 
In our context, leadership-driven AI adoption mandates can lead to ineffective AI integration when organizational readiness does not align with leadership expectations \cite{mayer2025superagency}, which can cause structural strain on teams who are pushed to adopt AI with ineffective results. Even when there is organizational readiness, developers' \textit{workload} can increase if they are assigned more tasks or expected to complete tasks at a faster pace by management because of AI adoption. Workload can also increase if developers have to spend additional time debugging or fixing AI errors \cite{harness2025}. Past work has shown that these complementary factors, organizational pressure and workload, are significantly associated with elevated work stress \cite{rosca2021job, scholze2024job}. We therefore hypothesize: \textbf{H1: Job demands related to GenAI adoption are associated with developer burnout.}

\textbf{Job Resources} refer to the physical, social, or organizational aspects of work that help individuals achieve work goals, reduce job demands, or stimulate personal growth and development \cite{demerouti2001job, bakker2017job}. It includes the two formative components: (1) Autonomy and (2) Organizational support as its constituents, both of which have been shown to buffer the adverse effects of job demands on burnout \cite{bakker2007job}.

\textit{Autonomy} in our context reflects the degree of control developers retain over how they integrate AI tools into their work, and is a well-known factor in improving employee well-being, motivation, and performance when navigating changing work environments \cite{ravn2022team}. \textit{Organizational support}, such as providing learning resources, can enable developers to maximize the value of AI \cite{houck2025space}. Other, non-AI-related research has consistently shown that autonomy and organizational support are important aspects of job resources to reduce burnout and foster engagement in technology-driven work environments \cite{scholze2024job, rosca2021job}. Similarly, reviews of JD--R applications in leadership contexts emphasize autonomy as a central resource for mitigating strain and enhancing performance \cite{tummers2021leadership, bakker2004using}. Building on these insights, we hypothesize that: \textbf{H2: Job resources related to GenAI adoption are associated with developer burnout.}

\textbf{Burnout} is a state of exhaustion, cynicism, and reduced efficiency caused by prolonged stressors, including work pressures and intrinsic factors such as mental stress or job insecurity \cite{maslach2001job}. Organizational mandates for GenAI adoption can amplify job insecurity when routine development tasks are advertised as being automated or when management signals potential workforce reductions \cite{molino2020promotion}. Additionally, constant news of AI adoption as well as the breakneck speed of AI advancement can induce fear of deskilling \cite{crowston2025deskilling}. However, burnout, a complex social construct, can be mitigated when an individual is driven and excited by new advancements (in GenAI). Research shows that favorable technology beliefs (e.g., usefulness and trust) can reduce strain and mitigate negative outcomes \cite{davis1989perceived, venkatesh2012consumer}. Therefore, we hypothesize that \textbf{H3: Favorable perceptions of AI are associated with developer burnout.}

\textbf{Developer Characteristics.}  
The effects of job demands and access to resources are not uniform across an organization and are dependent on factors such as seniority, role, or organizational context \cite{demerouti2001job}. To examine whether these characteristics influence different constructs of the JD--R model, we hypothesize: \textbf{H4: Developer characteristics are associated with variation in: Burnout (H4a); Organizational Pressure (H4b);  Workload (H4c); Autonomy (H4d); and Learning Resources (H4e).}

In the upcoming sections, we first describe the survey design and data collection process in Section~\ref{sec:survey}. In Section~\ref{sec:RQ1}, we evaluate H1--H3 using Partial Least Squares--Structural Modeling (PLS-SEM) to understand the relationships among job demands, job resources, AI perceptions, and burnout (RQ1). Then, in Section~\ref{sec:RQ2}, we assess H4a--H4e through regression analyses of developers' characteristics (role, industry seniority, organization size) to assess variation in demands, resources, and burnout (RQ2). We then complement these quantitative analyses with qualitative reporting of open-ended responses to contextualize developers' experiences with GenAI adoption.

%% file: sections/3_Survey.tex
\section{Data Collection: Survey}

We collected data to test the above hypotheses through a survey of software professionals. Three researchers with experience in survey methodology and software engineering research collaboratively developed the survey instrument questions over the course of one month using the JD--R model as a guide \cite{demerouti2001job}. The constructs in the JD--R model are defined as \textit{latent constructs}: constructs that cannot be directly measured or observed, and instead are measured through a set of indicators (survey items). We adapted existing survey questions to measure the different constructs.

\textbf{Survey design.} The survey began with the university’s Institutional Review Board (IRB) approved consent form, followed by five sections comprising a total of 15 questions. See the complete questionnaire in the supplementary document Section 1 \cite{supply}.

The first survey section collected participant characteristics, including \textit{gender}, \textit{professional background} (primary role in software development, years of industry experience, and seniority level); \textit{organizational context} (organization size), \textit{AI-usage experience} (frequency of AI use in software development work, adapted from \cite{usage2025}), and \textit{attitude towards AI adoption} measured by agreement with the statement: ``I feel positive about AI usage''.

The following section, related to \textit{Job Demands}, asked questions about organizational pressure and workload resulting from AI adoption. We measured  \emph{organizational pressure} by asking participants to rate 4 statements on a 5-point Likert scale (1 = strongly disagree, 5 = strongly agree; N/A if not applicable). These questions were adapted from work on technology adoption and job design \cite{brown2002really, morgeson2006work}, which showed that organizational mandates, collaborative expectations, and heightened responsibility can generate pressure. Some sample questions are: ``I must also use AI to get my work done as my co-workers rely on AI outputs''; ``My performance evaluations depend on how effectively I use AI tools''. \emph{Workload} questions also used the same 5-point Likert scale and asked participants to rate 4 statements. These statements were adapted from \citet{cao2024jd} work that used JD--R to examine the impact of COVID-19 on workload. Two example questions are: ``I often have to do more work than I can do well''; ``My job requires me to work harder''.

The third survey section was about \textit{Job Resources} and included questions related to autonomy and learning resources. We measured \textit{autonomy} through questions, adapted from \citet{clausen2019danish}, about participants' perceived discretion and decision-making power in their development work, using a 5-point Likert scale (from ``To a very small extent'' to ``To a very large extent''; N/A if not applicable). Some example questions are: ``Do you have any influence on how you carry out your work tasks?''; ``Do you have sufficient authority to deal with the responsibilities you have in your work?''. For measuring availability of \textit{resources for learning AI}, we asked participants to rate the statement: ``My organization gives me the time, training, and resources I need to master AI'' on a 5-point Likert scale (1 = strongly disagree, 5 = strongly agree; N/A if not applicable). The survey then followed with an open-ended question about the AI training provided by participants’ employers, adapted from \cite{oliverwyman2024ai}.

The last survey section measured \emph{burnout} through participants’ perceptions through a set of 4 statements which participants rated on a 5-point Likert scale (1 = strongly disagree, 5 = strongly agree; N/A if not applicable). These questions were adapted from \citet{trinkenreich2024predicting} to capture established dimensions of burnout: workload manageability and exhaustion, loss of interest as an indicator of cynicism, and insecurity about one's future as a reflection of reduced efficacy. Sample questions include: ``I feel mentally and physically exhausted from work''; ``I feel secure in my job and confident about my future in this company''.

The survey ended with an open-ended question for sharing any final thoughts or experiences on how adopting AI tools has impacted participants' work. Participants could share their email if they wanted to participate in a raffle for a \$50 Amazon gift card. The survey took between 5-8 minutes to complete.

\textbf{Sandbox and Pilot Survey.} Before distributing the survey, we sandboxed the survey with five participants who had research experience in SE. The survey was iteratively refined through the sandbox sessions until no further concerns remained. We piloted the survey with two collaborators: one from a multinational technology corporation and the other from a U.S. national laboratory. Both found the survey straightforward, easy to understand, and relevant.

\textbf{Participant Recruitment.} To ensure broad coverage across technical backgrounds, seniority levels, organization sizes, and project domains, we recruited software developers from 56 Open Source Software (OSS) communities spanning diverse domains, including organizational repositories (e.g., Microsoft, Google, Netflix, Red Hat, IBM) as well as widely used open-source infrastructure, cloud, and AI projects (e.g., Kubernetes, Hugging Face). We also reached out to data science-oriented communities (e.g., Python, TensorFlow) and professional groups. We \textit{intentionally used a broad recruitment strategy} to reflect the evolving nature of software development, which increasingly involves not only traditional software engineers but also data scientists, machine learning practitioners, and AI developers working together in collaborative software development environments \cite{feng2025domains}.

In accordance with ethical data collection guidelines, we sent email invitations detailing the survey's purpose, along with a consent form that contained information on the survey's voluntary nature, potential risks, and data protection measures. All participants provided informed consent, and responses were anonymized in compliance with the General Data Protection Regulation (GDPR) and our IRB approval. The survey was available for two weeks. This recruitment approach aligns with prior SE research \cite{feng2022case, feng2025multifaceted, feng2025domains}.

\input{Tables/demographics_distribution}

\textbf{Participant Distribution.} We received 688 responses. After excluding invalid responses, the final dataset comprised 442 participants. Table \ref{tab:demo_ref_style} summarizes the participants' characteristics. The most frequently reported primary area of work was full-stack software development ($n = 160$, 36.20\%). Of the 442 participants, 398 identified as men (90.05\%) and 44 as gender minorities \footnote{The term \textit{gender minorities} includes participants who identified as non-binary or gender diverse, selected “prefer not to say,” or reported multiple gender identities.} (9.95\%). More than half the participants reported working at large or extra-large companies (56.79\%), and the majority ($n = 360$) reported having six or more years of industry experience.

%% file: Tables/demographics_distribution.tex
\begin{table}[!htbp]
\vspace{-2mm}
\caption{Characteristics of Participants (N=442).}
\centering
\resizebox{2.5 in}{!}{
\begin{tabular}{lrr}
\hline
\multicolumn{1}{c}{\textbf{Attribute}} &
\multicolumn{1}{c}{\textbf{N}} &
\multicolumn{1}{c}{\textbf{Percentage}} \\[2pt]
\hline

\multicolumn{3}{c}{\textbf{Gender}} \\ \hline
Man & 398 & 90.05\% \\
\rowcolor[HTML]{EFEFEF} Gender Minorities & 44 & 9.95\%  \\\hline

\multicolumn{3}{c}{\textbf{Role}} \\\hline
Full Stack Software Development & 160 & 36.20\% \\
\rowcolor[HTML]{EFEFEF} Backend Development & 74 & 16.74\% \\
Data/ML Engineering & 66 & 14.93\% \\
\rowcolor[HTML]{EFEFEF} Frontend Development & 22 & 4.98\% \\
Project Management & 19 & 4.30\% \\
\rowcolor[HTML]{EFEFEF} System Architecture & 16 & 3.62\% \\
DevOps & 15 & 3.39\% \\
\rowcolor[HTML]{EFEFEF} Technical Writing & 7 & 1.58\% \\
Security Engineering & 7 & 1.58\% \\
\rowcolor[HTML]{EFEFEF} Quality Assurance & 2 & 0.45\% \\
 System Administration & 1 & 0.23\% \\
\rowcolor[HTML]{EFEFEF} Other & 53 & 11.99\% \\ \hline

\multicolumn{3}{c}{\textbf{Organization Size}} \\ \hline
S (fewer than 50 employees) & 140 & 31.67\% \\
\rowcolor[HTML]{EFEFEF} M (between 50 and 249 employees) & 51 & 11.54\% \\
L (between 250 and 4999 employees) & 104 & 23.53\% \\
\rowcolor[HTML]{EFEFEF} XL (5000 employees and more) & 147 & 33.26\% \\\hline

\multicolumn{3}{c}{\textbf{Experience}} \\\hline
1-5 years & 82 & 18.55\%\\
\rowcolor[HTML]{EFEFEF} 6-10 years & 96 & 21.72\% \\
11-20 years & 153 & 34.62\% \\
\rowcolor[HTML]{EFEFEF} Over 20 years & 111 & 25.11\% \\\hline

\end{tabular}}
\Description{Characteristics of Participants (N=442).}
\label{tab:demo_ref_style}
\end{table}

%% file: sections/4_RQ1_PLS.tex
\section{Impact of AI Adoption on Burnout (RQ1)}
\label{sec:RQ1}
\begin{figure}[!htbp]
    \vspace{-3mm}
    \centering
    \includegraphics[width=\columnwidth]{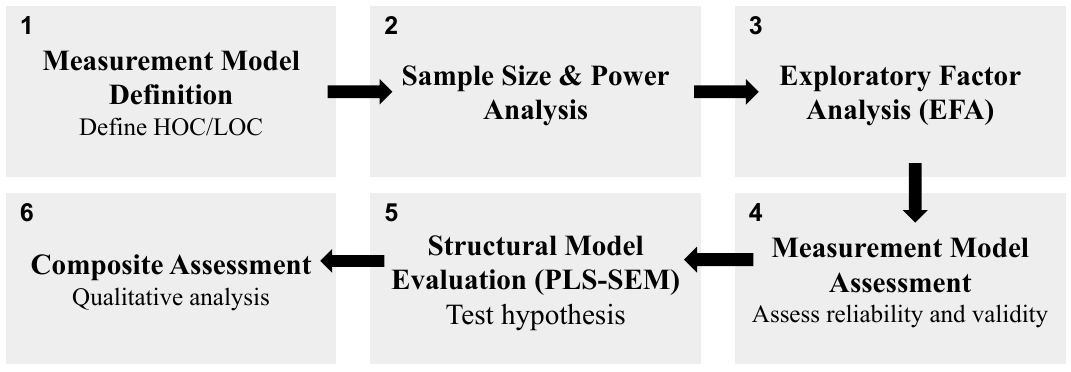}
    \caption{Overview of the analysis process for addressing RQ1.}
    \Description{Overview of the analysis process for addressing RQ1: 1. Measurement model definition 2. Sample size and power analysis 3. Exploratory factor analysis (EFA) 4. Measurement model assessment 5. Structural model evaluation (PLS-SEM) 6. Composite assessment}
    \label{fig:RQ1}
    \vspace{-3mm}
\end{figure}

To address \textbf{RQ1}, we followed a concurrent embedded mixed-methods approach \cite{greene2007mixed, morse2016mixed} (as shown in Figure \ref{fig:RQ1}). We first leveraged Partial Least Squares--Structural Equation Modeling (PLS--SEM) \cite{ringle2015structural} to analyze the survey data to evaluate hypotheses H1 to H3 through steps: \textbf{\textcircled{1}}- \textbf{\textcircled{5}}. We then conducted qualitative analysis on open-ended question responses to capture developers' perspectives on how AI adoption reshapes demands and resources (\textbf{\textcircled{6}}).

\subsection{Method}
\label{sec:RQ1method}

\textbf{PLS-SEM analysis}, a multivariate analysis technique, is our analysis of choice as it allows simultaneous testing of multiple relationships among constructs, while accounting for measurement error in modeling \textit{latent} constructs (e.g., Burnout) \cite{trinkenreich2024predicting}. It is also increasingly being used in empirical SE to investigate complex phenomena \cite{trinkenreich2025investigating, choudhuri2024far}. The JD--R theoretical foundation, as explained in Section \ref{sec:theory}, serves as the guiding lens for our analysis and includes the steps outlined in Figure~\ref{fig:RQ1}.

\textbf{\textcircled{1} Measurement Model}. We modeled \emph{Burnout} as a construct measured by four \textit{reflective} indicators from our survey items. These indicators capture different yet complementary facets of burnout: workload manageability (Burn1), emotional and physical exhaustion (Burn2), cynicism/disengagement (Burn3), and reduced efficacy/job insecurity (Burn4). In a \textit{reflective} measurement model, the indicators covary---e.g., when developers are emotionally or physically exhausted, they are less effective---and the construct (Burnout) is inferred from their shared variance.

\emph{Job Demands} was modeled as a \textit{formative} higher-order construct (HOC) comprising two lower-order constructs (LOCs)-- Organizational Pressure and Workload. We modeled these LOCs as formative indicators of Job Demands (HOC) for both theoretical and methodological reasons. First, consistent with the JD--R Literature, job demands represent a set of heterogeneous stressors that collectively increase strain, but need not covary, that is, they are not interchangeable reflections of a single latent trait. For example, a high workload does not necessarily imply high organizational pressure; yet, both independently contribute to burnout. Second, formative modeling allows us to account for the fact that removing one LOC (e.g., workload) would fundamentally alter the Job Demands construct, rather than simply reducing measurement reliability. This treatment aligns with prior applications of the JD--R model, which have modeled different types of demands as complementary contributors to stress and burnout \cite{schaufeli2004job}.

The LOC Organizational Pressure modeled as a latent construct measured by four \textit{reflective} indicators (Org1-Org4), which captured structural and managerial expectations tied to AI adoption (e.g., ``Management expects me to use AI''). Similarly, Workload was measured by four \textit{reflective} indicators (Work1-Work4) that capture the perception of intensified task requirements (e.g., ``I often have to do more work than I can handle well'').

\emph{Job Resources} was modeled as a \textit{formative} HOC composed of two reflectively measured LOCs-- Autonomy and Learning Resources. These two constructs are modeled as formative indicators of Job Resources (HOC), as they neither covary nor are they interchangeable (e.g., having autonomy does not imply getting AI training resources, and vice versa). This is in line with prior research, which has modeled job resources formatively when combining distinct categories such as social support, feedback, and autonomy \cite{schaufeli2004job}. Autonomy, a latent construct, was measured using three reflective indicators from our survey items assessing developers' influence over how they carry out their tasks (Auto1), their authority to handle responsibilities (Auto2), and their ability to make important decisions about their work (Auto3). Learning Resources was a single-item (Res) measure capturing whether developers had access to organizational support for acquiring AI-related knowledge and skills.

\emph{AI-perception} was modeled as a control variable and measured through a single item capturing developers' overall attitudes toward AI adoption, which may influence developer strain \cite{davis1989perceived, venkatesh2012consumer}. For the complete mapping of survey items to indicators of constructs, see Section 2 in the supplementary document \cite{supply}.

\textbf{\textcircled{2} Sample Size and Power Analysis.}  To ensure the adequacy of our sample size, we conducted a power analysis using the G*Power tool \cite{faul2009statistical}. We have three predictors of burnout in our model (Job Demands, Job Resources, and AI Perceptions as a control). Assuming a medium effect size ($f^2 = .15$), significance level ($\alpha = .05$), and desired statistical power ($1-\beta = .95$), the analysis indicated a minimum required sample size of 119. Our sample size of 442 far exceeds this requirement.

\textbf{\textcircled{3} Exploratory Factor Analysis (EFA)} was needed because, although the JD--R constructs have been validated in prior literature \cite{scholze2024job, rosca2021job, dorta2023s}, we had to adapt the survey items to our context of developer workflows and AI adoption. EFA allowed us to verify that the items clustered in ways that matched our intended constructs and that the underlying factor structure was consistent with our JD--R model \cite{fabrigar1999evaluating, worthington2006scale}. All EFAs were performed using JASP \cite{jasp2023}.

Following established guidance \cite{costello2005best, worthington2006scale}, we first confirmed that the data met the necessary assumptions for factor analysis (1) sampling adequacy (Kaiser--Meyer--Olkin $\geq .60$) \cite{hair2009multivariate}, and (2) sufficient intercorrelations among items (a significant Bartlett's test of sphericity) \cite{bartlett1950tests}. We then evaluated items against multiple criteria: (3) factor loadings ($\geq .50$ preferred, $.30$–$.50$ considered marginal) \cite{hair2009multivariate,hair2019use}, (4) cross-loadings ($< .30$) \cite{hair2019use}, (5) uniqueness ($< .60$) \cite{hair2009multivariate}, and (6) parallel analysis (keeping only the factors with eigenvalues larger than the corresponding average random eigenvalues) and inspection of the scree plot (keeping factors before the curve levels off) for final determination of factor retention \cite{howard2016review}. See Tables 2-4 in the supplementary document for the test results \cite{supply}.

The results supported a two-factor structure for Job Demands (Organizational Pressure and Workload), a one-factor structure for Job Resources (Autonomy), and a one-factor structure for Burnout. Because Learning Resources was measured with a single item, it was not included in the EFA \cite{hair2009multivariate, worthington2006scale}. Work1, Work2, Burn2, and Burn4 items had elevated uniqueness ($>.60$), but were retained because their factor loadings exceeded the recommended threshold of .50, which is considered a strong indicator of construct relevance in social science research \cite{costello2005best, fabrigar1999evaluating}.

\textbf{PLS-SEM analysis.} We then conducted PLS-SEM analysis using SmartPLS 4 \cite{ringle2015structural}. PLS-SEM involves two steps: (\textbf{\textcircled{4}}) evaluating the measurement model to empirically assess the relationships between the construct and its constituent indicators, and (\textbf{\textcircled{5}}) evaluating the structural model to test the hypothesized relationships among constructs. These results are reported in Section~\ref{sec:RQ1result}.

\textbf{\textcircled{6} Composite Assessment.} To better understand developers' experiences with AI adoption, we qualitatively analyzed participants' responses ($N = 221$) to the open-ended survey question: ``If you have any final thoughts or experiences on how adopting AI tools has impacted your work, please share them below.'' We used LOCs (e.g., Organizational Pressure, Workload, Autonomy, Learning Resources) and the control variable (AI-Perception) from our model as our codeset. Two authors independently coded 20\% of the responses, and compared their outputs using the Jaccard index, achieving 90\% agreement on inter-rater reliability (IRR) \cite{landis1977measurement}. The remaining responses were then divided between the two authors and coded individually using the stable codebook. See codebook in the supplementary document Table 15 \cite{supply}.

\subsection{Results for RQ1}
\label{sec:RQ1result}

\textbf{\textcircled{4} Measurement Model Assessment.}  
We used a two-stage evaluation to assess the (1) reflective measurement model and (2) formative measurement model as per \citet{becker2012hierarchical,sarstedt2019specify}.

\textbf{Stage 1: Reflective Measurement Model Evaluation.} We evaluated the reflective constructs (LOCs: Organizational Pressure, Workload, Autonomy, Learning Resources) and the endogenous construct (Burnout) by using tests for: (i) convergent validity, (ii) internal consistency reliability, (iii) discriminant validity, and (iv) collinearity among reflective indicators \cite{hair2019use, hair2020assessing}.

\begin{figure}[!htbp]
    \vspace{-3mm}
    \centering
    \includegraphics[width=\columnwidth]{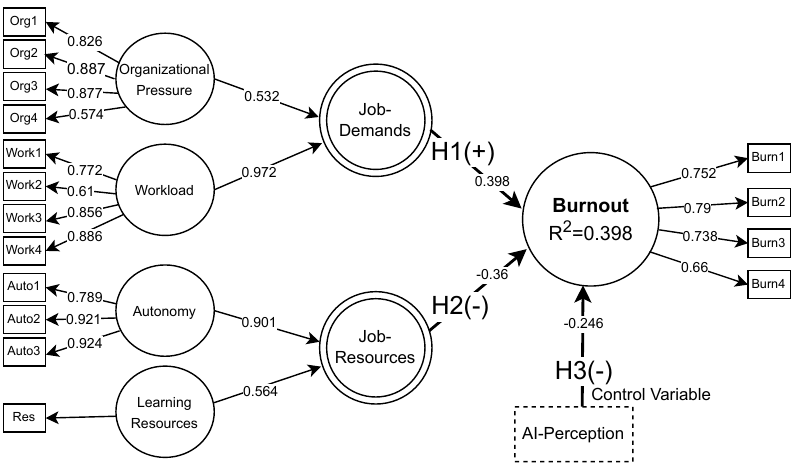}
    \caption{Outer loadings and path coefficients (p < 0.05 indicated by a full line; $N=442$). Higher order constructs (e.g. Job-Demands) are represented by ``double circles'' and have paths to their corresponding lower order constructs (e.g. Workload); a dashed rectangle represents ``Control Variable'' (e.g. AI-Perception).}
    \Description{Outer loadings and path coefficients (p < 0.05 indicated by a full line). Higher order constructs (e.g. Job-Demands) are represented by ``double circles'' and have paths to their corresponding lower order constructs (e.g. Workload); a dashed rectangle represents ``Control Variable'' (e.g. AI-Perception).}
    \label{fig:plsm}
    \vspace{-3mm}
\end{figure}

\emph{(i) Convergent validity} examines whether indicators intended to measure the same construct share sufficient variance \cite{kock2014advanced}. We assessed this using (a) outer loadings and (b) Average Variance Extracted (AVE) \cite{hair2019use}. (a) Outer loadings represent the bivariate relation between each indicator and its latent construct; values $\ge .70$ shows that the indicator shares at least 50\% of its variance with the construct. Indicators with outer loadings between $0.40$ -- $0.70$ can be removed when doing so improves construct-level reliability, or AVE, or if they are theoretically essential \cite{hair2020assessing}. In our model, as shown in Figure \ref{fig:plsm}, Work2 (0.61), Org4 (0.574), and Burn4 (0.66) fell within this range, but were retained because removing them did not improve construct-level reliability. All other indicators exceed the recommended threshold of 0.70. (b) AVE summarizes the proportion of common variance across indicators and should exceed $0.50$, indicating that the latent construct explains at least half of the variance in its indicators \cite{hair2019use}. Table~\ref{tab:reliability_cr_ave} shows that the AVEs meet the criteria.

\emph{(ii) Internal consistency reliability} determines whether indicators consistently reflect their underlying construct and is assessed using Cronbach's $\alpha$ and Composite Reliability (CR: $\rho_a$, $\rho_c$) tests \cite{russo2021pls}.  The acceptable range for both these values is between $0.70$ and $0.9$ \cite{hair2019use}. Table~\ref{tab:reliability_cr_ave} shows all LOCs and Burnout is in the acceptable range.

\input{Tables/construct_reliability_AVE}

\emph{(iii) Discriminant validity} determines whether each construct is empirically distinct from the others and includes three tests: (a) the Heterotrait-Monotrait (HTMT) ratio of correlations, where values below .90 indicate adequate discriminant validity; (b) indicator cross-loadings, where each item should load higher on its intended construct than on others; 
and (c) the Fornell-Larcker criterion, where the square root of each construct's AVE should be greater than its correlations with other constructs \cite{henseler2015new, hair2019use}. All constructs meet these criteria and are reported in Tables 7-9 in the supplementary document \cite{supply}.

\emph{(iv) Collinearity check of reflective indicators} ensures that indicators did not exhibit multicollinearity among themselves. We used the Variance Inflation Factor (VIF), where values below 3 to 5 are considered acceptable \cite{hair2019use}, and all indicators in our model satisfy this criterion (see Table 10 in the supplementary document \cite{supply}).

\textbf{Stage 2: Formative Measurement Model Evaluation.}  
We then evaluated the formative constructs (HOCs) following the assessment steps \cite{becker2012hierarchical,sarstedt2019specify}: (i) the significance of outer weights, (ii) outer loadings, and (iii) collinearity among the LOCs for each HOC.

\emph{(i) Outer weights} capture the unique contribution of each LOC to its HOC after controlling for the effects of the other LOCs. Significant outer weights ($p < .05$) indicate that the LOC adds explanatory power to the HOC beyond redundancy. For Job Demands, the weight of Workload is statistically significant ($p < .05$), while the weight of Organizational Pressure is marginal ($p = .064$). For Job Resources, the weights of both Autonomy and Learning Resources are statistically significant ($p < .05$).

\emph{(ii) Outer loadings} assess whether each LOC contributes meaningfully to its HOC by reflecting their bivariate correlation. Values above $0.50$ are generally considered acceptable evidence of relevance, even when outer weights are marginal \cite{sarstedt2019specify}. In our case, the outer loading of Organizational Pressure is $>0.5$ and therefore is retained, although it has marginally significant outer weights.

\emph{(iii) Test for collinearity among LOCs} ensures that each HOC's constituent LOCs do not exhibit multicollinearity. We used the VIF, and all values are well below the recommended cutoffs (below 3). See Tables 11-13 for Stage 2 analysis in the supplementary \cite{supply}.

\textbf{\textcircled{5} Structural Model Evaluation.}  We evaluated the structural model by (i) assessing collinearity among predictor constructs, (ii) testing the hypotheses, and (iii) examining the explanatory power of the model. Figure~\ref{fig:plsm} presents the model.

\emph{(i) Collinearity among predictor constructs} was assessed, and it was found that predictor constructs in the structural model (Job Demands, Job Resources, and AI Perceptions) do not exhibit multicollinearity by calculating VIFs. All VIF values are well below the threshold of 3 (see Table 14 in the supplementary document\cite{supply}).

\emph{(ii) Path coefficients} were estimated to identify the direct effects of one construct on another in the structural model. Path coefficients were assessed using nonparametric bootstrapping with 5,000 subsamples, which provides empirical standard errors, $p$-values, and confidence intervals for inference \cite{hair2019use}. Table~\ref{tab:path-coeff} reports the path coefficients ($\beta$), standard deviations (SD), 95\% confidence intervals, and $p$-values.

We found \textbf{empirical support for hypotheses H1 and H2}. As shown in Figure~\ref{fig:plsm}, \textbf{Job Demands} (i.e., organizational pressures and workload intensification) \textbf{are \textit{positively associated} with burnout} ($\beta = 0.398$, $p<.001$). In contrast, \textbf{Job Resources related to GenAI adoption} (i.e., autonomy and learning resources) \textbf{are \textit{negatively associated with burnout}} ($\beta = -0.360$, $p<.001$). For the LOCs of Job Demands, Workload has a stronger contribution, while for Job Resources, Autonomy has a stronger contribution. \textbf{Hypothesis H3 also has empirical support}. \textbf{AI perception is \textit{negatively associated with burnout}} ($\beta = -0.246$, $p<.001$), indicating that more favorable AI perceptions correspond to lower burnout.

\emph{(iii) Explanatory Power of Burnout} was assessed using the coefficient of determination ($R^2$). The $R^2$ value for Burnout is 0.398 (Studies in organizational and social science contexts note that $R^2$ values around $.30$ -- $.40$ represent moderate explanatory power, given the complexity of human and organizational behavior \cite{cohen2013statistical, chin1998partial}).

We assessed the model's external validity by assessing its predictive relevance using Stone-Geisser's $Q^2$ via blindfolding in SmartPLS. Blindfolding iteratively omits and predicts portions of the data (using an omission distance $D$) and computes $Q^2$ from prediction errors, where values above 0 indicate predictive relevance \cite{stone1974cross, hair2021partial}. In our case, the $Q^2$ value for Burnout is 0.373, indicating a large predictive relevance.

We assessed the model fit using the standardized root mean square residual (SRMR) \cite{russo2021pls}, with a value less than 0.08 (conservative) or 0.10 (more lenient) considered a good fit. Our models' SRMR meets the threshold. 


\input{Tables/path-coefficients}

\textbf{\textcircled{6} Composite Assessment.} To complement the quantitative findings, our qualitative analysis captures developers' perspectives on how AI adoption reshapes job demands and resources.

\emph{Organizational Pressure (Job Demands).} AI was \textit{``over-hyped''} [P228] and enforced as \textit{``a race to the bottom''} [P69], with \textit{``C-level staff \ldots carrying the banner''} [P177] and \textit{``we are forced to use AI so a metric can be reported to the management even if we don't need it''} [P169]. For many, the tools themselves were less problematic than the managerial narratives around them, \textit{``the tools are fine, what they've done to leadership expectations is absolutely awful''} [P187].

\emph{Workload (Job Demands).} Participants described how GenAI heightened their workload, often pushing them beyond sustainable limits. \textit{``let's say my car has a manufacturer's recommended optimal speed of 65 mph with the optimization criteria of engine longevity. But then everyone is using this cool new fuel additive, which boosts the engine and makes it regularly go at a seemingly effortless 100 mph. I feel like my brain is working at 100mph, instead of the optimal 60 mph''} [P384]. Beyond mental load, GenAI also introduced additional rework \textit{``more time is spent cleaning up after AI tools than if we were to\ldots write code from scratch''} [P259] and \textit{``I wasted more time tracking down subtle bugs''} [P14].

\emph{Autonomy (Job Resources).}
When developers retain the freedom to decide when and how to use AI, autonomy can act as a buffer against burnout. In some contexts, developers emphasized that adoption was optional and unobtrusive \textit{``Its use is not required\ldots work requirements and deliverables are unchanged''} [P5]; \textit{``I am\ldots free to continue my previous pre-AI behavior, sprinkling in AI use where I think it is beneficial''} [P175]. Elsewhere, the situation was reversed with organizational mandates narrowing choices: \textit{``We are forced to use AI, so a metric can be reported to the management''} [P169]. 
Such imposed choices led to inefficient workflows \textit{``Each code injection is a mini peer-review which breaks my chain of thought''} [P194] or constrained access to desired tools \textit{``I often feel restricted by the limited rollout of certain tools''} [P340].

\emph{Learning Resources (Job Resources).} 
\textit{``Blindly trusting AI outputs can cause more problems, so training is required to ensure staff do not get into the habit of using AI outputs without validation''} [P18]. 
Organizational mandates without providing sufficient scaffolding was a recurring problem: \textit{``[Company has] broad statements encouraging us to use AI to help us be more productive, but have never offered any training on how to do it''} [P375]. \textit{``AI is not a magic solution; the LLM approach needs strict guidance \& ideal examples to mimic''} [P92]. To further understand what organizational resources are actually provided and how developers access them, we analyzed open-ended survey responses in Section \ref{sec:RQ2} to identify what teams receive, in what forms, and with what gaps.

\emph{AI-Perception.}
Negative AI-perceptions aligned with heightened strain, review burden, and insecurity: \textit{``By my estimate, LLM-based tools have so far been a net negative\ldots\ I quickly wasted more time tracking down subtle bugs\ldots''} [P13]. \textit{``I feel like we're in the uncanny valley of when these tools will be truly useful\ldots\ and it is draining''} [P80], whereas, positive perceptions alleviated the strain: \textit{``My experience is generally positive. Agentic systems (I mostly use Cursor) allow me to offload tedious tasks and accelerate my research''} [P4].

%% file: Tables/construct_reliability_AVE.tex
\begin{table}[!htbp]
\vspace{-3mm}
\caption{Internal Consistency Reliability and Convergent Validity}
\centering
\resizebox{3in}{!}{
\begin{tabular}{lrrrr}
\hline
\multicolumn{1}{c}{} &
\multicolumn{1}{c}{\textbf{$\mathrm{Cronbach's}$  $\alpha$}} &
\multicolumn{1}{c}{\textbf{$\mathrm{CR}(\rho_a)$}} &
\multicolumn{1}{c}{\textbf{$\mathrm{CR}(\rho_c)$}} &
\multicolumn{1}{c}{$\mathrm{AVE}$} \\[2pt]
\hline
\rowcolor[HTML]{EFEFEF} Burnout & 0.718 & 0.724 & 0.825 & 0.542 \\
Organizational Pressure & 0.827 & 0.879 & 0.875 & 0.642 \\
\rowcolor[HTML]{EFEFEF} Workload & 0.801 & 0.845 & 0.866 & 0.621 \\
Autonomy & 0.856 & 0.893 & 0.911 & 0.775 \\
\hline

\multicolumn{5}{p{3.5in}}{\footnotesize Cronbach's $\alpha$ tends to underestimate reliability, whereas composite reliability ($\mathrm{CR}: \rho_c$) tends to overestimate it. The true reliability typically lies between these two estimates and is effectively captured by $\mathrm{CR}(\rho_a)$ \cite{russo2021pls}. \textit{Note: Learning Resources was measured with a single item and is therefore not included in this table, as reliability and AVE are not applicable for single-item construct.}}
\end{tabular}
}
\Description{Internal Consistency Reliability and Convergent Validity}
\label{tab:reliability_cr_ave}
\vspace{-3mm}
\end{table}


%% file: Tables/path-coefficients.tex
\begin{table}[!htbp]
\caption{Standardized path coefficients (B), standard deviations (SD), confidence intervals (95\% CI), p-values ($p$).}
\resizebox{3.2in}{!}{
\begin{tabular}{lllllll}
\hline
\multicolumn{1}{c}{} &
\multicolumn{1}{c}{$\mathrm{B}$} &
\multicolumn{1}{c}{$\mathrm{SD}$} &
\multicolumn{1}{c}{95\% $\mathrm{CI}$} &
\multicolumn{1}{c}{$\mathrm{\textit{p}}$} \\[2pt]
\hline
\rowcolor[HTML]{EFEFEF} H1 Job-Demands $\rightarrow$ Burnout & 0.398 &  0.044 & (0.313	, 0.483) & 0.000  \\
H2 Job-Resources $\rightarrow$ Burnout & -0.36 & 0.042 & (-0.445, -0.278) & 0.000  \\
\rowcolor[HTML]{EFEFEF} H3 AI-Perception $\rightarrow$ Burnout & -0.246 &
0.048 & (-0.334, -0.148) &  0.000 \\
\hline

\end{tabular}
}
\Description{Standardized path coefficients, standard deviations, confidence intervals (95\% CI), p-values.}
\label{tab:path-coeff}
\vspace{-3mm}
\end{table}

%% file: sections/5_RQ2_logistic.tex
\section{Influence of Developer Characteristics (RQ2)} 
\label{sec:RQ2}

To answer \textbf{RQ2}, we continue to follow the mixed-methods concurrent embedded approach \cite{greene2007mixed, morse2016mixed}. We first used ordinary least squares (OLS) regression models \cite{fox2015applied} to assess the relationship between developer characteristics and job demands, resources, and burnout. We then qualitatively analyzed open-ended responses on job resources to contextualize our analysis results, surfacing how learning resources are (or are not) made available in practice.

\subsection{Data Analysis}

We built five separate regression models, where each of the LOCs — Organizational Pressure, Workload, Autonomy, Learning Resources, and Burnout — was the dependent variable (on a 1–5 Likert scale), and the independent variables were the developers’ characteristics, including developer role, organization size, and industry experience.

\emph{Dependent Variables.} 
To create a composite score of the dependent variables that were measured using multiple indicators (e.g., burnout includes four indicators), we use the outer weights of each indicator estimated in the measurement model (see Table 6 in \cite{supply}). For example, 
$\text{Burnout} = \sum_{i=1}^{4} \text{Burn}_i \cdot w_i$, where $\text{Burn}_i$ are the items (Burn1--Burn4) and $w_i$ are their corresponding outer weights. This approach is standard in survey-based research, where responses to multiple Likert-scale items are aggregated to form a single interval-level measure \cite{sullivan2013analyzing, chikezie2025measuring}.

\emph{Independent Variables.} All five regression models used the same set of developer characteristics as independent variables. \textit{Organization size} was coded as 4-level  ordinal variable: small (S = 1), medium (M = 2), large (L = 3), and extra-large (XL = 4). \textit{Industry experience} was treated as an ordinal variable representing industry seniority, coded as: less than 5 years = 1, 6–10 years = 2, 11–20 years = 3, and more than 20 years = 4. Finally, \textit{Role} was coded as a binary variable (frequent coder = 1, non–frequent coder = 0); participants whose primary area of focus was software development (e.g., Backend Development) were considered frequent coders, and those with other focus areas (e.g., Technical Writing) were considered non-frequent coders.

\emph{Model Assessments.} We assessed the models' explanatory power by calculating the $R^2$ and adjusted $R^2$ values. In social science research, $R^2$ values in the range of 0.10–0.30 are often considered acceptable given the complexity of human behavior \cite{ozili2023acceptable}. Our results fall within this range.

To address Type I error inflation from multiple comparisons, we applied Benjamini-Hochberg adjustment to the $p$-values \cite{benjamini1995controlling}. 
Table \ref{tab:linear-regression} shows the regression analysis results with adjusted $p$-values.

\subsection{Results for RQ2}

\input{Tables/linear-regression}

Results of the OLS regression coefficients are presented in Table \ref{tab:linear-regression}. Each cell reports the estimated effect ($\beta$) of developer characteristics (Coder, Organization Size, and Industry Seniority) on the dependent variables (Burnout, Organizational Pressure, Workload, Autonomy, and Learning Resources). \textbf{We do not have empirical support for H4a}, indicating that in our dataset, \textbf{burnout is a broadly experienced phenomenon regardless of developers’ characteristics}.

\textbf{Job Demands.} \textbf{\emph{Organizational Pressure} is positively associated with coding role} ($\beta=0.51, p<0.05$) \textbf{and organization size} ($\beta=0.41, p<0.001$) \textbf{and are statistically significant.} These results are consistent with prior work showing that larger organizations impose adoption mandates, thereby heightening the pressure for developers \cite{bakker2007job, scholze2023digital}. 
AI-based coding support has seen significant advances and is touted for improving developer productivity \cite{peng2023impact, paradis2025much}; thus, developers in coding-intensive roles likely encounter greater strain with AI adoption \cite{harness2025, ravn2022team}. Industry seniority is not statistically significant. Thus, (\textbf{H4b) is supported only for role and organization size.}

Turning to the second job demand dependent variable, \textbf{\emph{workload}, the regression model reveals no significant associations with developer characteristics, showing no empirical support for (H4c}).

\textbf{Job Resources.} \textbf{We have empirical support for (H4d) organization size and industry seniority}. \textbf{Organization size is negatively associated with \emph{autonomy}} ($\beta=-0.13, p<0.001$), indicating that developers in larger organizations perceive less discretion and decision-making power. This is consistent with research showing that bureaucratic structures and standardized processes often constrain individual influence \cite{tummers2021leadership}. \textbf{Industry seniority is positively associated with autonomy} ($\beta=0.18, p<0.001$), suggesting that senior developers have greater say over tasks and responsibilities, echoing findings that experience confers both expertise and organizational credibility \cite{clausen2019danish, hackman1976motivation}. Prior research has shown that large organizations often introduce new tools such as AI through top-down mandates that restrict individual discretion \cite{scholze2023digital}. However, senior developers are better positioned to shape adoption practices by leveraging their expertise and credibility \cite{clausen2019danish}.

For \emph{learning resources}, we find that organization size and industry seniority show statistically significant associations. \textbf{Developers in larger organizations report greater access to learning resources} ($\beta = 0.28$, $p < .001$), \textbf{and more experienced developers similarly report higher availability} ($\beta = 0.14$, $p < .05$). \textbf{We have empirical support for (H4e) for organization size and industry seniority}.

\input{Tables/Resources_Types}

\textbf{Understanding Resource in Practice.} Our regression analysis shows that developer characteristics are associated with job demands and job resources. 
However, external conditions that result in organizational pressure, such as the need for AI adoption or structural conditions such as organization size or industry seniority, cannot be altered in the short term. 
But learning resources are something that organizations and developers can intervene more directly (e.g., through training, playbooks, or guardrails). To better understand what forms of support are being provided and how developers access them, we included an open-ended survey question on organizational supports (Section~\ref{sec:survey}).

Two authors first conducted open coding to identify preliminary categories of resources. They then iteratively refined these codes to resolve discrepancies and collaboratively construct a final codebook through negotiated agreement \cite{garrison2006revisiting, forman2007qualitative}. Using the codebook, two authors independently coded 20\% of the responses and compared results to calculate IRR using the Jaccard index\cite{landis1977measurement}, achieving 92\% agreement.

The final codeset comprises six categories of resource types, ranging from formal training programs to tool support. Table~\ref{tab:resources} provides a summary of these categories and their frequencies.

Among the 361 open-ended responses, 22.4\% indicates that organizations provided no meaningful support for learning resources. The most frequently reported forms of support are internal training (30.2\%) and access to training artifacts (19.04\%). However, when mapping responses to organization size and developer seniority, we observe that resource availability may depend on organizational scale, which in turn supports our regression analysis showing that organization size is positively associated with learning resources.

Another notable proportion (15.23\%) is self-study, where participants rely on their own time and financial resources to learn. From the distribution mapping, we observe that self-study is more common in small organizations. In terms of seniority, more senior developers tend to rely on self-study, suggesting a stronger sense of responsibility or motivation to continue learning -- \textit{``they only provide very basic prompt engineering, for that reason I enrolled in some university courses myself''} [P436].

%% file: Tables/linear-regression.tex
\begin{table}[!htbp]
\vspace{-3mm}
\caption{Estimated regression coefficients ($\beta$) for the effects of developer characteristics (Coder, Organization Size (Size), Industry Seniority (Years)) on dependent constructs (Burnout, Organizational Pressure, Workload, Autonomy, Learning Resources; $N=442$)}
\resizebox{3in}{!}{
\begin{tabular}{llll}
\hline
\textbf{} & \textbf{Coder}  & \multicolumn{1}{c}{\textbf{Size}} & \textbf{Years}  \\
\hline
\rowcolor[HTML]{EFEFEF} Burnout    & -0.06   & 0.08 & -0.12 \\
Organizational Pressure & \textbf{0.51*} & \textbf{0.41***}  & -0.06 \\
\rowcolor[HTML]{EFEFEF} Workload   & -0.02 & 0.01 & -0.02\\
Autonomy & -0.06 & \textbf{-0.13***}   & \textbf{0.18***}\\
\rowcolor[HTML]{EFEFEF} Learning Resources & 0.33 & \textbf{0.28***} & \textbf{0.14*}\\
\hline
\multicolumn{4}{p{3.05in}}{\footnotesize Values marked with $^{\ast}$ are statistically significant using BH-corrected $p$-values: $(p<.05)^{\ast}$, $(p<.01)^{\ast\ast}$, $(p<.001)^{\ast\ast\ast}$.}

\end{tabular}
}
\Description{Estimated regression coefficients ($\beta$) for the effects of developer characteristics (Coder, Organization Size (Size), Industry Seniority (Years)) on dependent constructs (Burnout, Organizational Pressure, Workload, Autonomy, Learning Resources; $N=442$)}
\label{tab:linear-regression}
\vspace{-4mm}
\end{table}

%% file: Tables/Resources_Types.tex
\begin{table*}[!htbp]
\caption{Learning resource types reported by developers. Six categories of organizational support were identified from 361 open-ended responses. The table also displays the frequency of each category along with distributions by company size and developer seniority (five ordinal levels each).}
\resizebox{0.9\textwidth}{!}{ 
\begin{tabular}{llllll}
\textbf{Code} & \textbf{Definition} & \textbf{Example} & \textbf{Frequency} & \textbf{Size} & \textbf{Seniority} \\
\rowcolor[HTML]{EFEFEF} 
\begin{tabular}[c]{@{}l@{}}Internal Training / Workshops / \\
Bootcamps\end{tabular} 
& \begin{tabular}[c]{@{}l@{}}Company-organized learning with a\\scheduled format (live or recorded)\end{tabular} 
& \textit{``In-person workshops"} [P129] 
& \barChart{30.2} 
& \miniHistogram[63]{11,10,33,63}
& \miniHistogram[40]{19,33,40,25}\\
\begin{tabular}[c]{@{}l@{}}Access to training artifacts\end{tabular} 
& \begin{tabular}[c]{@{}l@{}}Informal, self-paced learning materials—videos,\\docs, tutorials, wikis, tip sheets and online courses\end{tabular} 
& \begin{tabular}[c]{@{}l@{}} \textit{``Written articles, links to videos, examples}\\ \textit{and time to experiment with the tools"} [P222] \end{tabular} 
& \barChart{19.04} 
& \miniHistogram[46]{4,8,17,46}
& \miniHistogram[25]{18,13,25,18}\\
\rowcolor[HTML]{EFEFEF} 
\begin{tabular}[c]{@{}l@{}}Self-study (any resources/ MOOC/ \\LinkedIn Learning)\end{tabular} 
& \begin{tabular}[c]{@{}l@{}}Learning outside working hours or\\“on my own time,” and money\end{tabular} 
& \textit{``Self-guided learning during work time"} [P15]
& \barChart{15.23} 
& \miniHistogram[23]{23,4,19,14}
& \miniHistogram[25]{4,11,20,25}\\
\begin{tabular}[c]{@{}l@{}}Tool supports\end{tabular} 
& Paid subscription 
& \textit{``Paid Cursor subscription"} [P34]
& \barChart{7.87} 
& \miniHistogram[10]{10,4,8,9}
& \miniHistogram[17]{7,4,17,3}\\
\rowcolor[HTML]{EFEFEF} 
\begin{tabular}[c]{@{}l@{}}Peer Mentoring /Learning \& \\Communities\end{tabular} 
& \begin{tabular}[c]{@{}l@{}}Peer-to-peer knowledge sharing or\\ mentoring \end{tabular}
& \begin{tabular}[c]{@{}l@{}} \textit{``Mostly peer support and mentoring"} [P61] \end{tabular}
& \barChart{6.6} 
& \miniHistogram[11]{8,2,11,5}
& \miniHistogram[13]{3,5,13,5}\\
\begin{tabular}[c]{@{}l@{}}Budget\end{tabular} 
& \begin{tabular}[c]{@{}l@{}}Financial support earmarked for learning—\\ course fees, conference/workshop registrations\end{tabular} 
& \textit{``Budget to pay courses online"} [P54]
& \barChart{4.57} 
& \miniHistogram[6]{5,6,3,4} 
& \miniHistogram[7]{2,4,7,5}\\
\rowcolor[HTML]{EFEFEF} 
\begin{tabular}[c]{@{}l@{}}Organization provided formal \\Courses \&  Certifications\end{tabular} 
& \begin{tabular}[c]{@{}l@{}}External or formally credentialed learning with\\ assessments/certificates\end{tabular} 
& \begin{tabular}[c]{@{}l@{}} \textit{``We have certification programs which}\\ \textit{are mandatory before getting access to} \\ \textit{the AI tools"} [P7] \end{tabular} 
& \barChart{2.03} 
& \miniHistogram[4]{1,0,3,4}
& \miniHistogram[3]{2,1,3,2}\\
\end{tabular}
}
\Description{Learning resource types reported by developers. Six categories of organizational support were identified from 361 open-ended responses. The table also displays the frequency of each category along with distributions by company size and developer seniority (five ordinal levels each).}
\label{tab:resources}
\end{table*}

%% file: sections/6_limitation.tex
\section{Limitations}

In empirical studies of heterogeneous domains, theory is crucial for guiding analysis and preventing purely descriptive interpretations \cite{stol2018abc}. We adopted the JD–R model as our theoretical lens because it enables systematic analysis of the interplay between job demands, job resources, and practitioner well-being, and it has been widely applied in organizational and SE contexts \cite{bakker2017job, trinkenreich2024predicting}. We acknowledge that some aspects of AI adoption may fall outside the scope of JD–R; however, we included an open-ended survey question to ask participants to share additional thoughts about AI adoption. Our qualitative analysis of these responses revealed no additional themes beyond those covered by our survey items.

Our hypotheses in this study propose associations between different constructs in the JD-R model rather than causal relationships, as the present study is a cross-sectional sample study \cite{stol2018abc}. For example, participants who were already experiencing higher burnout (e.g., due to upcoming deadlines) might adopt AI differently compared to those under lighter workloads, which may introduce response bias. Other confounding factors, such as trust in AI systems or perceived productivity, may have influenced results but were not directly measured. Therefore, our results should be interpreted as a theoretical starting point, guiding future studies to explore these contextual influences.

When answering RQ1, we employed PLS–SEM. We acknowledge that there are other methods, such as CB–SEM, but we selected PLS–SEM because our JD–R model includes both formative and reflective constructs, which PLS–SEM is better suited to analyze \cite{hair2011pls}. Moreover, PLS–SEM offers stronger explanatory power, making it appropriate for investigating developer well-being in the emerging context of GenAI adoption \cite{hair2011pls}.

While no single sample can capture the entire global software workforce, our sample of 442 software practitioners from 56 organizations is comparable in size to, or larger than, many empirical studies in software engineering \cite{russo2024navigating, trinkenreich2023belong}. More importantly, our participants represent a wide range of company sizes, job roles, and levels of experience, which supports diversity in perspectives and provides a suitable starting point for understanding the associations presented in our model. 

%% file: sections/7_discussion.tex
\section{Concluding Remarks}

\subsection{AI Induced Workflow Shifts.}

The introduction of GenAI to software engineering has generated extraordinary excitement, with industry narratives promising unprecedented productivity gains. Yet, as with earlier waves of technological enthusiasm, the promise of relief is overlaid with forms of work pressure and strain \cite{khan2023excessive}.

\textbf{Shift 1: From euphoria to stress.} AI hype like ``AI can do everything'' and ``make everyone faster'', has translated into mandates for higher output and expectation. \textit{``I'm expected to do more now. Although AI is helpful, it's not that helpful that it would make that big of a difference. At the end, I am doing more work than before''} [P191]. AI adoption often is not relief but work intensification with little room to slow down. This pressure is amplified by organizational narratives that assume AI will succeed everywhere and for everyone.  \textit{``Everything AI-related is suddenly a Priority-1 Emergency...it feels like I am constantly in a Priority-1 Emergency''} [P291].

\textbf{Shift 2: From apprenticeship to private copy--paste.} Pre GenAI, newcomers' learning progression was coupled with social learning practices such as pair programming, code reviews, mentoring, and public Q\&A ecosystems \cite{feng2022case, feng2025multifaceted}. Today, however, practices like Vibe coding and pair programming with AI have emerged as the new trend \cite{peng2023impact, ray2025review}. While these tools may accelerate certain tasks, they also narrow social networks, reduce mentoring opportunities, and erode the ``resources'' that historically supported early-career growth. Participants raised concern that this shift could delay skill acquisition, \textit{``I fear that juniors will take much longer to get to an experience level that is comparable to what devs with 10+''} [P202].

\textbf{Shift 3: From productivity gains to hidden collaboration costs.} Collaboration friction has long been a fixture in software development \cite{feng2025domains}, GenAI use can increase the workload for colleagues, adding to the friction: \textit{``reviewing LLM-generated content such as code and docs wastes time, coworkers are (accidentally but carelessly) sabotaging our work by creating [more] work''} [P134]. Another participant said: \textit{``AI generates many test cases, including verbose ones, which makes code-reviewing more time-consuming''} [P25]. While authors may save time by pasting in AI outputs, reviewers inherit the burden of checking quality, correcting errors, and editing down verbosity; \textit{``I spend a far greater share of my time editing their writing and checking the validity of their assertions''} [P374].

\subsection{Implications For:}
\textbf{Workload Design.} Studies abound that extol AI for its ability to improve productivity \cite{peng2023impact, paradis2025much}. However, adoption of AI can create additional downstream workload, as reported by our participants, which can erase the productivity gains. When paired with misplaced management expectations of ever-increasing output, it causes burnout. In fact, one industry survey found that 77\% of AI users reported increased workload because of AI adoption \cite{unleash2024_ai_workload_upwork}. 
This is because, while AI reduces some types of work (e.g., automated code or test generation), it creates other types (e.g., in-depth verification, debugging AI-generated code). AI is shifting the type of work from creation towards curation and oversight, which is cognitively demanding and requires in-depth technical expertise, but is harder to track.

To sustain productivity without causing burnout, organizations need to redesign their performance metrics.  Key Performance Indicator (KPI) should explicitly account for both the new capabilities AI affords and the verification costs it introduces. Without such recalibration of organizational goals, AI adoption can put developers in an unsustainable cycle of doing more, faster, under shrinking timelines.

\textbf{Workforce Development.} Our analysis showed that larger organizations and senior developers were more likely to access AI-related resources. This does not necessarily mean that resources were entirely unavailable to others, but rather that senior developers were more inclined and had more opportunity to seek them out. Table~\ref{tab:resources} further indicates that seniors reported higher engagement in self-study. One possible explanation is that juniors, having just entered the workforce, face heightened expectations and tighter deadlines, leaving little time to engage in deliberate learning. Under this pressure, the immediate goal becomes task completion, which encourages copy--paste practices: \textit{``Junior developers don't know how to use it and they blindly apply it without thinking it through before submitting''} [P246]. The long-term danger is a generational fault line, in which today’s experienced developers retire and leave behind a cohort of ``GenAI-native'' developers with weaker foundations in core knowledge of software design and architecture, debugging, and testing.

Organizations need to treat developers' learning as an investment, especially for newcomers. Rather than relying on AI as a shortcut to higher throughput, managers should allocate explicit time for juniors to reflect on AI outputs, review them with peers, and connect with mentors. The rapid pace of AI advancement means that workforce development cannot be treated as a one-time training event; instead, it needs to be scaffolded as a continuous, career-long learning objective. Developers need access to structured upskilling opportunities as well as the time to upskill, so that all developers--seniors and juniors--have equitable access to AI-related training. Without such scaffolding, disparities in experience risk hardening into structural inequalities in capability.

The responsibility of workforce development does not solely rest with industry. If junior developers are entering the workforce with a predisposition for superficial AI usage, then educators need to address this gap earlier on. Universities need to update their learning objectives to explicitly encourage students to be responsible AI consumers: teaching students both the foundational SE principles as well as how to reflect and critically think about AI responses. Embedding AI literacy throughout the curriculum while reinforcing core SE skills is essential to avoid the erosion of fundamentals. Academic–industry partnerships, through micro-training contextualized to bespoke industry problems, co-developed case studies, and internships, can help align educational preparation with the realities of a rapidly changing workplace.

\textbf{Team Workflow.} AI adoption can cause collaboration friction because of the perception of others producing large amounts of low-quality work: \textit{``Coworkers use LLMs to generate a large amount of low-quality content (pull requests, documents) \textemdash{} bordering on ‘spam’.''} [P134]; \textit{``This week I reviewed documentation clearly written by AI. It needed a lot of work… I was spending more time reviewing and re-reviewing than the author had spent writing it.''} [P124]. Such dynamics shift team effort from creation to quality assurance--efforts neither evenly distributed nor evenly recognized--causing team friction, which in turn reduces social support and peer learning opportunities, exacerbating burnout.

To avoid these pitfalls, organizations should accompany AI rollouts with explicit workflow adaptations. Teams can integrate practices such as: (i) declare provenance by noting when and how AI was used, (ii) curate AI-augmented outputs by editing and de-verbosing content, (iii) reduce workload by segmenting changes and providing evidence of quality control through tests, linters, and security scan outputs, and (iv) improve workload equity by rotating reviewer responsibilities and updating KPIs to include these efforts.

In \textbf{conclusion}, our findings show that current hype cycles and mandate-driven adoption of GenAI can intensify job demands, unevenly distribute resources, and contribute to developer burnout. Yet, history reminds us that each technological wave, from assembly lines to the automobile to the internet, has reshaped not only what we produce, but how we work. Some practices will fade, and new ones will emerge as the new norm. As participant eloquently summarized: \textit{``I think it has a lot of potential to be used as a carefully curated tool in one's tool belt, like a chisel or a hammer can be for a woodworker but I think trying to force it's adoption without proper training to enforce such a mindset is going to lead to a large brain drain in developers which will have negative long term ramifications''} [P117]. Sustainable development workflows, therefore, depend not only on adopting AI tools but on building the human and organizational capacity to manage their downstream effects.